\documentclass[twocolumn,showpacs,preprintnumbers,amsmath,amssymb,prl]{revtex4}

 \usepackage[dvips]{graphicx}% Include figure files
 \usepackage{dcolumn}% Align table columns on decimal point
 \usepackage{bm}% bold math

\begin{document}

%\preprint{Submitted to {\bf PRL}}

\title{High-pressure molecular phases of solid carbon dioxide}

\author{S. A. Bonev, F. Gygi, T. Ogitsu, and G. Galli  }
\affiliation{Lawrence Livermore National Laboratory,
              University of California, Livermore, California 94550}
\email{bonev1@llnl.gov} 
              
\date{\today}

\begin{abstract}
We present a theoretical study of solid carbon dioxide (CO$_2$) 
up to 50~GPa and 1500~K using first-principles calculations.
In this pressure-temperature range, interpretations of recent 
experiments have suggested the existence of CO$_2$
phases which are intermediate between molecular and covalent-bonded solids.
We reexamine the concept of intermediate phases in the CO$_2$ phase
diagram and propose instead  molecular structures, which provide 
an excellent agreement with measurements.
\end{abstract}


\maketitle

%--- MATH----
\newcommand{\co}{CO$_2$}
\newcommand{\sio}{SiO$_2$}
%-------------

%%%%%%%%%%%%%%%%%%%%%%%%%%%%%%%%%%%%%%%%%%%%%%%%%%%%%%%%%%%%%%%%%%%%%%%%%%
%
%\section{Introduction}
%
%%%%%%%%%%%%%%%%%%%%%%%%%%%%%%%%%%%%%%%%%%%%%%%%%%%%%%%%%%%%%%%%%%%%%%%%%%
Understanding the evolution of the bonding properties of molecular 
crystals as a function of pressure is a fundamental question in 
condensed matter physics. Recently, this question has received a widespread 
attention due to key progress in experimental techniques
and hence, the availability of new data for a number of
molecular solids \cite{hem00}. 
In the case of CO$_2$, the interest 
has been further intensified due to its importance for 
planetary science and technical applications.
However, despite several experimental
investigations of the {\co} high-pressure phases, the changes
of intra- and inter-molecular bonds, as well as of electronic and 
vibrational properties, as a function of pressure 
are not yet well understood. In addition, large portions of the 
{\co} phase diagram are unexplored from a theoretical standpoint.

\begin{figure}[b!]
  \hspace*{-5mm}
  \includegraphics[height=0.19\textheight,width=0.4\textwidth,clip]{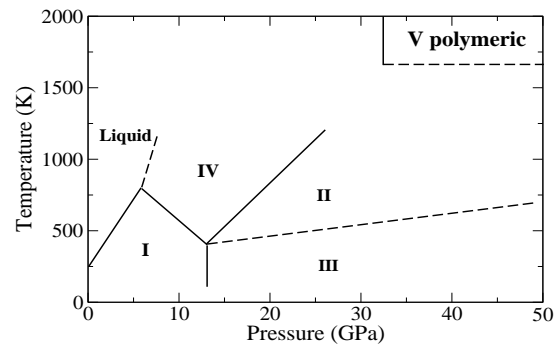}
  \caption{\label{fig_pdo} Phase diagram of {\co} according to 
           Ref~\onlinecite{iyo01}, where \co-III ($Cmca$) is reported 
           to be metastable.}
\end{figure}

At low pressure ($P$) and temperature ($T$), 
{\co} condenses as a molecular solid in the 
cubic $Pa3$ structure (known as dry ice, phase I), 
which is characterized by strong double
bonds (C=O distance of 1.16~{\AA}) and rather weak inter-molecular 
interactions \cite{kko34}.
Recently, a high-$P$ and high-$T$ phase of CO$_2$ (phase V) 
has been discovered \cite{ycg99,iyc99,scc99,dts00,hab00}, 
which is completely different from the molecular solid.
It has a polymeric quartz-like structure, and 
a very low compressibility (the experimentally derived bulk modulus,
$B_0$, is 365~GPa). Following this exciting discovery, 
two additional new phases
have been reported \cite{iyo01,yic01,ykc02} experimentally 
(II and IV in Fig.~\ref{fig_pdo}),
both of which were credited with unusual properties;
their structures were described as intermediate
between that of molecular and covalently bonded crystals. 

In this Letter, we present a study of the CO$_2$ phase diagram 
based on first-principles density functional theory (DFT) calculations.
Our results challenge the interpretation of \co-II, III, 
and  IV as exhibiting dramatic differences in the nature of the
molecular bonding with respect to the low-pressure molecular crystal.
In place of the previously proposed structures for \co-II and IV,
we suggest new, molecular ones. We demonstrate
that in addition to being stable, our newly 
proposed structures give results in excellent
agreement with measurements and provide a consistent explanation 
of experimental observations.   

At ambient $T$, \co-I undergoes a pressure-induced transformation 
to the orthorhombic $Cmca$  symmetry  
(phase III)  between 12 and 22~GPa 
\cite{ycg99,liu84,han85,ket88,ays94,lho95,oje98}.
The exact nature of this transition  and the structure of phase III  
are still unsettled; experimental studies have suggested a region of 
co-existence  between \co-I and III \cite{ays94} , and also an 
intermediate distorted low-$T$ phase \cite{oje98}. Furthermore, it was 
reported \cite{iyo01} that the I to III transition is strongly 
temperature dependent, becoming abrupt above 400~K, and that
\co-III may actually be metastable.
%---------------------------------------------------------------------------
%---------------------------------------------------------------------------
The two newly discovered phases II and IV  
are believed to be thermodynamically stable and quenchable. 
It was proposed that \co-II has the $P4_2/mnm$ symmetry
with elongated C=O bonds, 1.33~{\AA},  
reduced inter-molecular distances below 2.38~{\AA} 
(hence named associated or dimeric phase), 
and $B_0 = 131$~GPa \cite{iyo01,ykc02}.
\co-IV was described as $Pbcn$, with  C=O bonds as long
as 1.53~{\AA}, and bent (O=C=O angle of $160^\circ$) and strongly (dipole) 
interacting molecules~\cite{yic01}.
Such remarkable properties would imply that CO$_2$ is losing
its molecular character in a ``gradual'' way, a concept promoted
as a key for understanding its phase diagram in analogy with solids like
N$_2$O and SiO$_2$. 
In addition, \co-III, despite having a stable C=O bond,
was also assigned a $B_0$ as high as $87$~GPa \cite{iyo01}.
The stability of the proposed bent and associated phases,
and the unusual bulk properties of \co-II, III, and IV have not been 
investigated theoretically up to date.

%%%%%%%%%%%%%%%%%%%%%%%%%%%%%%%%%%%%%%%%%%%%%%%%%%%%%%%%%%%%%%%%%%%%%%%%%%
%
%\section{Computational method}
%
%%%%%%%%%%%%%%%%%%%%%%%%%%%%%%%%%%%%%%%%%%%%%%%%%%%%%%%%%%%%%%%%%%%%%%%%%%
We performed a series of first-principles calculations, including full 
structural optimizations, phonon spectra, 
and free energies, in order to study the 
stability and properties of the phases proposed experimentally up to 50~GPa 
and 1500~K. The DFT calculations were carried out within the 
Perdew-Burke-Ernzerhof (PBE) \cite{pbe96} generalized gradient
approximation (GGA) using the ABINIT code \cite{abini}, which 
implements plane-wave basis sets \cite{pwmtd}.

%%%%%%%%%%%%%%%%%%%%%%%%%%%%%%%%%%%%%%%%%%%%%%%%%%%%%%%%%%%%%%%%%%%%%%%%%%
%
%\section{Results and discussion}
%
%%%%%%%%%%%%%%%%%%%%%%%%%%%%%%%%%%%%%%%%%%%%%%%%%%%%%%%%%%%%%%%%%%%%%%%%%%
First, we have examined the stability of the bent and dimeric structures
for phases II and IV. Starting from the previously proposed $P4_2/mnm$ and
$Pbcn$ with elongated molecular bonds, we carried {\em full} 
structural optimizations at various pressures up to 50 GPa. 
In all cases, upon relaxing the atomic coordinates 
the C=O bond lengths decreased by about 15\% and 30\% for $P4_2/mnm$ 
and $Pbcn$ respectively, to become comparable to the free {\co} 
molecule \cite{mcbnd}. In the case of $Pbcn$,
the O=C=O angle also straightened to $180^\circ$. The energy differences
between the theoretically stable molecular structures and the associated 
and bent phases are respectively more than 3 and 6~eV per molecule.
These are energy scales corresponding to the breaking of a covalent bond. 
They are beyond the errors of the GGA and present a strong evidence
that the previously proposed bent and dimeric structures are not stable.
In order to further investigate this issue, we examine below published
experimental data for \co-II, III, and IV, and demonstrate that they 
can be explained in terms of two {\em stable} 
structures, $Cmca$ and $P4_2/mnm$, which are {\em strictly molecular}.

\begin{figure}[tb]
  \hspace*{-5mm}
  \includegraphics[height=0.25\textheight,width=0.45\textwidth,clip]{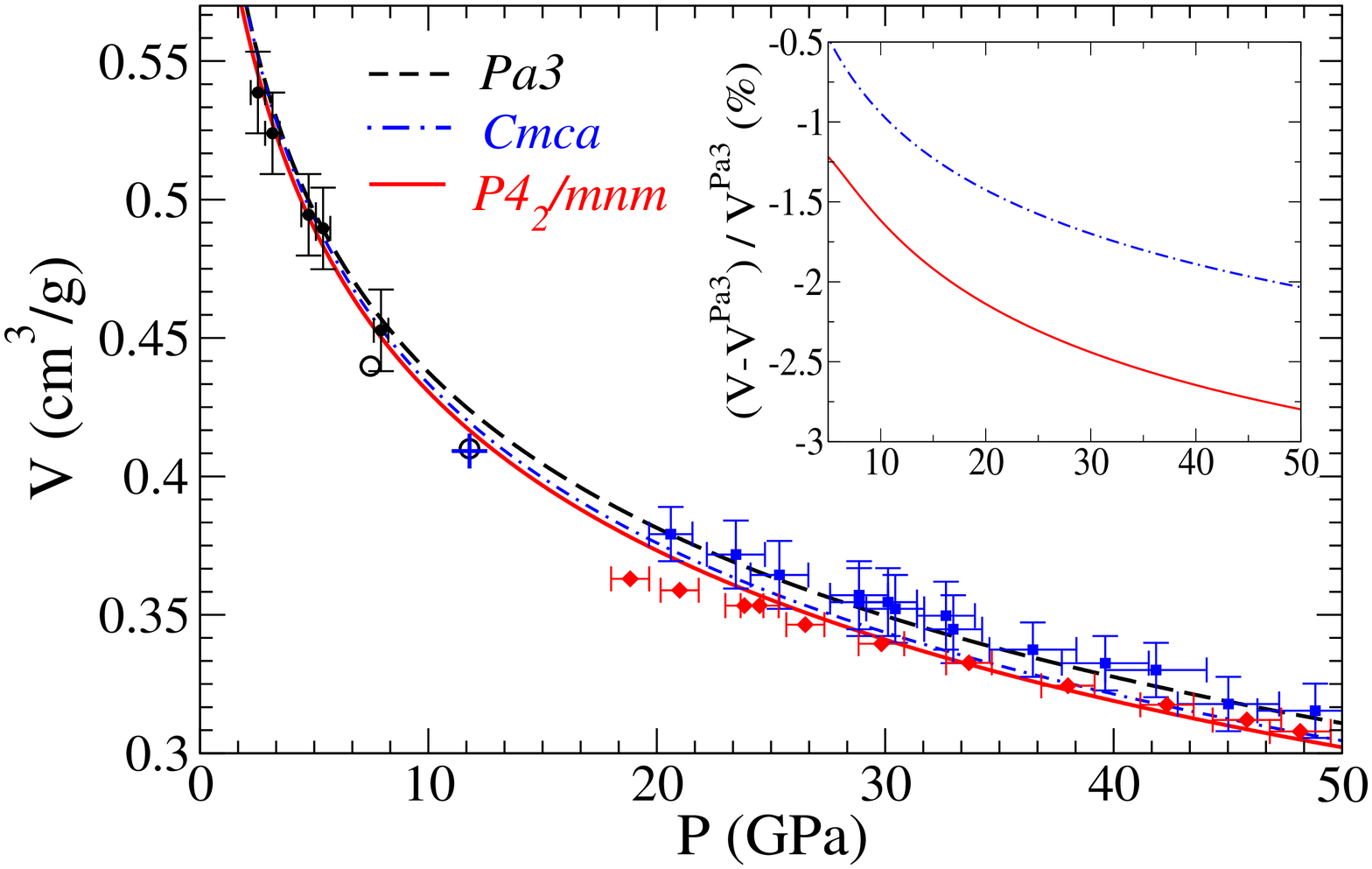}
  \caption{\label{fig_eos} 
           Pressure-volume dependence of selected {\co} structures:
           $Pa3$ (black), $Cmca$ (blue), and $P4_2/mnm$ (red).
           The solid lines are {\it ab initio} calculations; 
           the points indicate 
           experimental data from Ref.~\onlinecite{ays94} (no error bars),
           Ref.~\onlinecite{ycg99} ($Pa3$ and $Cmca$), and 
           Ref.~\onlinecite{ykc02} ($P4_2/mnm$). The inset shows calculated 
           reduction in volume of $P4_2/mnm$ and $Cmca$ relative to  $Pa3$.} 
\end{figure}

Computed equation of state (EOS) at $T=0$ for the three molecular 
structures $Pa3$, $Cmca$, and $P4_2/mnm$ are reported in Fig.~\ref{fig_eos}, 
along with the available experimental data. 
In all three cases, the calculated EOS agrees very well with the 
measured one; the computed $PV$ curves fall within the 
experimental uncertainties. We find about 2\%  relative volume
reduction from $Pa3$ to $Cmca$ \cite{cmcad} above 30~GPa and another 
0.5\% from $Cmca$ to $P4_2/mnm$ (see insert in Fig.~\ref{fig_eos}).
Yoo et al. reported a $\sim$ 5-7\% volume decrease associated with the 
III to II transition; however, they also observed lattice strain in 
their phase III samples \cite{iyo01}, which may explain the difference 
between their results and ours. Furthermore, the $PV$ curve of
$Cmca$ comes close to and eventually merges with that
of $Pa3$ below 20~GPa in agreement with Aoki et al. \cite{ays94}.  
Thus, the measured EOS data are reproduced well despite the lack of 
molecular association in our optimized $P4_2/mnm$. 
The computed EOS parameters are 
summarized in Table~\ref{tbl_eos} together with values 
derived form experiment \cite{ktm91}. The agreement is good for
$Pa3$, but there is an order of magnitude difference for the values
of $B_0$ of $Cmca$ and $P4_2/mnm$. Since we are using the same EOS fit as 
Ref.~\onlinecite{ykc02}, and we in fact agree very well with the 
{\em direct} $PV$ measurements, the large difference 
likely comes from extrapolating the experimental data,
spread above 20~GPa, down to ambient pressure.

%---------------------------------------------------------------------------
%---------------------------------------------------------------------------

%--------------------------------------------------------------------
\begin{table}[b]
\caption{\label{tbl_eos}
         Equation of state parameters for selected {\co} phases.
         The theoretical data is well fitted to a third-order 
         Birch-Murnaghan equation of state.}
\begin{ruledtabular}
\begin{tabular}{lcccc}
Structure & $V_0$ (cm$^3$/g) & $B_0$(GPa)& $B_0'$ & Ref. \\
\hline
%-------------------------------------------------
%
$Pa3$ & 0.714 &2.93 & 7.8 &\onlinecite{liu84} \\
$Pa3$ & 0.726 & 3.21  & 8.10 & this study \\
\hline
$Cmca$ & 0.450 & 87 & 3.3 &\onlinecite{ykc02}\\
$Cmca$ & 0.725 & 3.53 & 7.12  &this study \\
\hline
$P4_2/mnm$ & 0.408 & 131  & 2.1 &\onlinecite{ykc02}\\
$P4_2/mnm$ & 0.701 & 4.37 & 6.66  &this study \\
%-------------------------------------------------
\end{tabular}
\end{ruledtabular}
\end{table}
%--------------------------------------------------------------------

\begin{figure}[t]
  \hspace*{-5mm}
  \includegraphics[height=0.27\textheight,width=0.42\textwidth,clip]{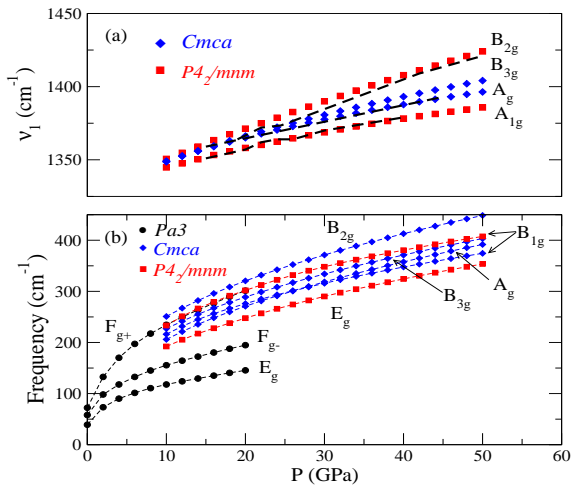}
  \caption{\label{fig_phon} Computed Raman-active modes (solid symbols) 
           of the theoretically stable structures.
	   (a) Symmetric stretching vibron. 
            The dashed lines indicate experimental data from
            Ref.~\onlinecite{iyo01}; they are from a Fermi resonance band, 
            and have been shifted by 52~cm$^{-1}$ for comparison.
           (b) External modes.} 
\end{figure}

The published experimental evidence invoked to 
propose that \co-II  exhibits the properties of a
dimeric polymorph consists of: ({\it i}) a large splitting of the internal
symmetric stretching mode, $\nu_1$ \cite{iyo01}, ({\it ii}) a 
broad librational mode identified as B$_{1g}$ \cite{iyo01,ykc02}, 
and ({\it iii}) powder x-ray diffraction measurements \cite{ykc02}. 
We carried out calculations of the vibron spectrum \cite{phonm} of
{\em molecular} $P4_2/mnm$ and $Cmca$ as a function of $P$. 
A plot of $\nu_1$ in Fig.~\ref{fig_phon}a shows that the
measured splitting, as well as the relative values for 
$P4_2/mnm$ and $Cmca$, are reproduced remarkably well by the theoretical 
structures. Yoo {\it et al.} assumed that the large vibron splitting 
observed in \co-II is an evidence for decreased  inter-molecular distances;
our results indicate that the crystal field in the molecular $P4_2/mnm$ 
is sufficient to explain the splitting. One should note that phonons are
calculated as a second derivative of the energy; therefore 
the agreement with experimental frequencies within meV, as
found here, is a strong indication that computed total energies
and forces are extremely accurate. 
 
The computed pressure dependence of the Raman-active external modes of 
the three theoretical structures is shown in Fig.~\ref{fig_phon}b. 
The agreement with experiment \cite{oje98} is again good, though 
our values are consistently slightly lower than the measured ones. 
The experimental frequencies of the two Raman-active modes of 
\co-II at 19~GPa  are about 260 and 320 cm$^{-1}$ - the
former being a broad peak - and were classified as B$_{1g}$ and E$_g$ 
respectively \cite{iyo01,ykc02}. The broad mode was previously 
associated with dynamical disorder in the lattice.
The molecular $P4_2/mnm$ has modes with frequencies 
245 and 300 cm$^{-1}$ at this $P$; however it is the degenerate  
E$_g$ which has the lower frequency, and therefore corresponds to 
the experimentally observed broad peak. Its broadening can 
be explained by invoking a lifting of the E$_g$ degeneracy 
when the tetragonal  cell is deformed into an orthorhombic one 
(resulting in the $Pnnm$ symmetry). Such a distortion is consistent with 
the x-ray data reported in Ref.~\onlinecite{ykc02} and
we estimate that between 20 and 30~GPa there is 10~cm$^{-1}$ 
splitting for every 1\% modifiaction of $a$ and $b$.

%---------------------------------------------------------------------------
%---------------------------------------------------------------------------

We now turn our attention to the x-ray analysis.
In $P4_2/mnm$, the atomic positions 
are C(2a) at [0,0,0], and O(4f) at [$x$,$x$,0]. The associated
phase, according to Ref.~\onlinecite{ykc02}, corresponds to $x=0.2732$, while
our stable structure corresponds to $x=0.23075$, i.e. 
$x\approx 0.25 \pm \delta$ 
for both structures, where $\delta = 0.02$. Among the observed 
diffraction peaks, only the (101) and (211) reflections depend on the 
sign of $\delta$; their measured intensities 
relative to the calculated intensities of the same reflections  
are in ratios of 2.3 and 0.3 for the associated, and 0.4 and 1.8 for the
molecular structures respectively\cite{yoopr} .
We therefore conclude that at present the 
diffraction measurements are insufficient to distinguish between 
the proposed dimeric and the theoretical molecular $P4_2/mnm$ structures.

%---------------------------------------------------------------------------
\begin{figure}[b!]
  \hspace*{-5mm}
  \includegraphics[height=0.32\textheight,width=0.42\textwidth,clip]{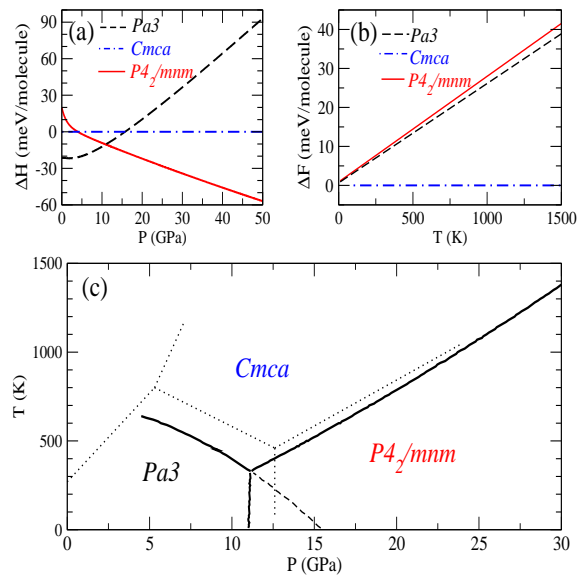}
  \caption{\label{fig_stab} (a) Enthalpy versus $P$
           relative to the $Cmca$ phase at $T=0$. 
	   (b) Phonon free energy versus $T$
           relative to the $Cmca$ structure at 16~GPa; the reported 
           differences do not vary strongly with $P$.
           (c) Relative stability of selected {\co} structures
           from Gibbs free energy comparison; here $P$
           includes phonon contributions. The dotted lines indicate the
           experimental phase diagram constraints from Ref.~\onlinecite{iyo01} 
          (see Fig.~\ref{fig_pdo}).}  
\end{figure}
%---------------------------------------------------------------------------

Finally, we examine the relative stability of all considered
crystal structures as a function of both $P$ and $T$. 
A plot of enthalpies showing that 
$Cmca$ is never thermodynamically stable at low $T$
is reported in Fig.~\ref{fig_stab}a.
This finding confirms a previous observation by Iota and Yoo \cite{iyc99}
that the orthorhombic symmetry obtained during ambient $T$
compression of \co-I is metastable; the reason why the system reverts to it is
the low kinetic barrier associated with the rotation of the molecules
from their alignment in $Pa3$ to that in $Cmca$. 
In additional support of this conclusion, we have performed structural
optimizations starting from the $Pbca$ structure, and relaxing the atomic
coordinates. The $Pbca$ lattice is obtained from $Pa3$ by deforming the
cubic cell; further rotations of the molecules in the $y$-$z$ plane 
then leads to $Cmca$. In all such calculations above 15~GPa, 
the {\co} molecules
quickly align in the $y$-$z$ plane, thus ending up in the $Cmca$, 
rather than in the energetically favorable $P4_2/mnm$ phase. This happens 
regardless of whether the unit cell is tetragonal or 
orthorhombic; the deformation of  $P4_2/mnm$ to the 
orthorhombic $Pnnm$ costs only about 
1~meV per molecule for 1\% distortion of $a$ and $b$.

The kinetic barrier between $Cmca$ and $P4_2/mnm$ can be overcome by heating 
the system above ambient $T$. Interestingly, we find that upon further
heating at constant $P$ there is a {\em thermodynamic} phase 
boundary above which the orthorhombic phase is now {\em stable}. 
The phonon free energies, computed as described in Ref.~\onlinecite{phonm}, 
are plotted in Fig.~\ref{fig_stab}b.
The transition to $Cmca$ is entropy driven 
and is mainly due to a soft acoustic phonon mode. Its physical origin 
is in the specific geometry of $Cmca$, allowing for a relatively unobstructed 
shear-like motion parallel to the $y$ axis; we have confirmed this 
by sampling the potential surface with frozen-phonon calculations. 

The computed stability regions for $Pa3$, $P4_2/mnm$, and $Cmca$ 
are shown in Fig.~\ref{fig_stab}c. Since  
we are comparing the relative stability of a {\em limited} number of 
structures, 
we cannot firmly conclude that 
Fig.~\ref{fig_stab}c represents the actual phase diagram of {\co}. 
However, the almost perfect matching of the stability region of $Cmca$ 
with that
of \co-IV makes the identification between the two phases rather compelling.
We note that a phase IV of {\co} was first suggested by
Olijnyk and Jephcoat \cite{oje98} based on the observed multiplicity of 
external Raman-active modes
between 10 and 20~GPa, at ambient $T$. Their data is consistent
with the Raman spectrum of $Pa3$, $Cmca$, and $Pbca$. 
As we discussed above, $Pbca$ is intermediate
between $Pa3$ and $Cmca$, and may be present close to the boundary of 
stability between these two phases, which extends from 5 to 16~GPa in our 
calculations.
In addition, $Cmca$ has a soft (below 100 cm$^{-1}$ at $\sim$ 20~GPa)
zone-boundary libron, B$_{1g}$,  tilting the molecules away from the 
$y$-$z$ plane, which crosses two of the acoustic modes.
Thus, when excited at high $T$, $Cmca$ could actually turn into a peculiar 
distorted phase. 
Ref.~\onlinecite{oje98} observed a similar distorted phase with  a 
weak (otherwise forbidden) Raman activity of the bending 
internal mode, which they related to the presence of pressure inhomogeneities
in the sample. We find such an explanation of the Raman activity 
observed by Yoo {\it et al.} \cite{yic01} around 650~cm$^{-1}$
more plausible than  the existence of bent molecules.

%%%%%%%%%%%%%%%%%%%%%%%%%%%%%%%%%%%%%%%%%%%%%%%%%%%%%%%%%%%%%%%%%%%%%%%%%%
%
%\section{Conclusion}
%
%%%%%%%%%%%%%%%%%%%%%%%%%%%%%%%%%%%%%%%%%%%%%%%%%%%%%%%%%%%%%%%%%%%%%%%%%%

In conclusion, we have presented a  new interpretation of 
experimental data recently obtained for \co-II, III, and IV. 
Our {\it ab inito} calculations
identify \co-II as a {\em molecular} structure  with the $P4_2/mnm$ symmetry.
We have also elucidated the high-$T$ behavior of molecular {\co}, and have 
found the $Cmca$
symmetry to be preferable due to its relatively large entropy. Based on
our findings, we propose that measurements of the high-$T$ 
phase IV be interpreted along the lines of 
an orthorhombic, and possibly distorted, phase. 
Finally, we note that the external Raman modes of several {\co} molecular
structures are very close in frequency (see Fig.~\ref{fig_phon}), 
and therefore may not be sufficient to discriminate between different
phases.

%%%%%%%%%%%%%%%%%%%%%%%%%%%%%%%%%%%%%%%%%%%%%%%%%%%%%%%%%%%%%%%%%%%%%%%%%%
%
%\section{Acknowledgments}
%
%%%%%%%%%%%%%%%%%%%%%%%%%%%%%%%%%%%%%%%%%%%%%%%%%%%%%%%%%%%%%%%%%%%%%%%%%%

We thank C.S. Yoo, J. Park, and  V. Iota for useful discussions.
This work was performed under the auspices of the U.S. Dept. of 
Energy at the University of California/LLNL under contract no. W-7405-Eng-48.

%%%%%%%%%%%%%%%%%%%%%%%%%%%%%%%%%%%%%%%%%%%%%%%%%%%%%%%%%%%%%%%%%%%%%%%%%%%
%

%
%%%%%%%%%%%%%%%%%%%%%%%%%%%%%%%%%%%%%%%%%%%%%%%%%%%%%%%%%%%%%%%%%%%%%%%%%%%


\begin{references}
\bibitem{mcbnd} In the case of $P4_2/mnm$ , the bond length is 1.154~{\AA} 
                at 28~GPa, and eventually reduces to 1.148~{\AA} at 50~GPa.
\end{references}
\end{document}